\documentclass[twocolumn,aps,prl]{revtex4}
\usepackage{exscale}                  
\usepackage{amsmath}
\usepackage{amsfonts}
\usepackage{amssymb,amscd}
\usepackage[dvips]{epsfig}                   
\usepackage{array}
\usepackage[usenames]{color}
\usepackage{graphicx}
\usepackage{bm}

\usepackage{graphics}
\usepackage{graphicx}
\usepackage{dcolumn}

\newcommand{\ar}{\arrowvert}
\newcommand{\ra}{\rangle}

\newcommand{\da}{\dagger}
\newcommand{\ov}{\overline}
\newcommand{\cd}{\! \cdot \!}
\newcommand{\be}{\begin{equation}}
\newcommand{\ee}{\end{equation}}
\newcommand{\bea}{\begin{eqnarray}}
\newcommand{\eea}{\end{eqnarray}}
\newcommand{\ba}{\begin{eqnarray}}
\newcommand{\ea}{\end{eqnarray}}
\begin{document}

\title{Heavy Quark Fluorescence}
\author{Juan M. Torres-Rincon and Felipe J. Llanes-Estrada}
\affiliation{Dept. F\'isica Te\'orica I, Universidad Complutense de Madrid, 28040 Madrid, Spain}

\begin{abstract}
Heavy hadrons containing heavy quarks (for example, $\Upsilon$-mesons) feature a scale separation between the heavy quark mass (about $4.5\ \textrm{GeV}$ for the $b$-quark) and the QCD scale (about $0.3\ \textrm{GeV}$) that controls effective masses of lighter constituents.
Therefore, as in ordinary molecules, the deexcitation of the lighter, faster degrees of freedom leaves the velocity distribution of the heavy quarks unchanged, populating the available decay channels in qualitatively predictable ways. Automatically an application of the Franck-Condon principle of molecular physics explains several puzzling results of $\Upsilon(5S)$ decays as measured by the Belle collaboration, such as the high rate of $B_s^* \overline{B}_s^*$ versus $B_s^*\overline{B}_s$ production, the strength of three-body $B^*\overline{B} \pi$ decays, or the dip in $B$ momentum shown in these decays. We argue that the data is showing the first Sturm-Liouville zero of the $\Upsilon(5S)$ quantum mechanical squared wavefunction, and providing evidence for a largely $b\overline{b}$ composition of this meson.
\end{abstract}

\maketitle

The Franck-Condon p‌rinciple~\cite{FranckCondon} is essential in understanding fluorescence in the decay of excited molecules. It states that, during an electronic transition, the electron orbital relaxes to the ground state in a time too short for the nuclei to react. Hence the wavefunction associated to the motion of the nuclei remains the same after the transition. Since the ground--state wavefunctions of this motion in the two electronic adiabatic potentials (before and after the transition) are unequal, the nuclei remain in a superposition of excited--states, resulting in secondary transitions (efficiently degrading excitation energy).

While the Born-Oppenheimer approximation from molecular physics had already been studied for decades in the context of particle physics, we have only recently understood~\cite{General:2006ed} the relevance of the Franck-Condon principle here.
The key is its usefulness to analyze open flavor decays of heavy hadrons.
Such decays are characterized by the heavy quarks from the initial hadron streaming away in separate hadrons. We take as example the newly analyzed $\Upsilon (10860)\to B\bar{B}$, $ B\bar{B}\pi \dots $ decays.

The momentum of the heavy quark inside the heavy hadron, $k_b$, follows a quantum-mechanical probability distribution that depends on the wavefunction of the state. This wavefunction accepts a decomposition in an appropriately chosen quark-gluon basis, for example
$\ar \Upsilon \ra = \alpha_0 \ar b\bar{b} \ra + \alpha_1 \ar b \bar{b} g \ra + 
\alpha_2 \ar b\bar{b} q\bar{q}\ra+\alpha_3 \ar ggg \ra+\dots $ \\
One of the salient problems of hadron physics nowadays is to determine the functions $\alpha_i$, whose size characterizes whether a state is ``mostly a quarkonium'' or otherwise.
\begin{figure}[htbp]
\begin{centering}
\includegraphics[width=6.6cm]{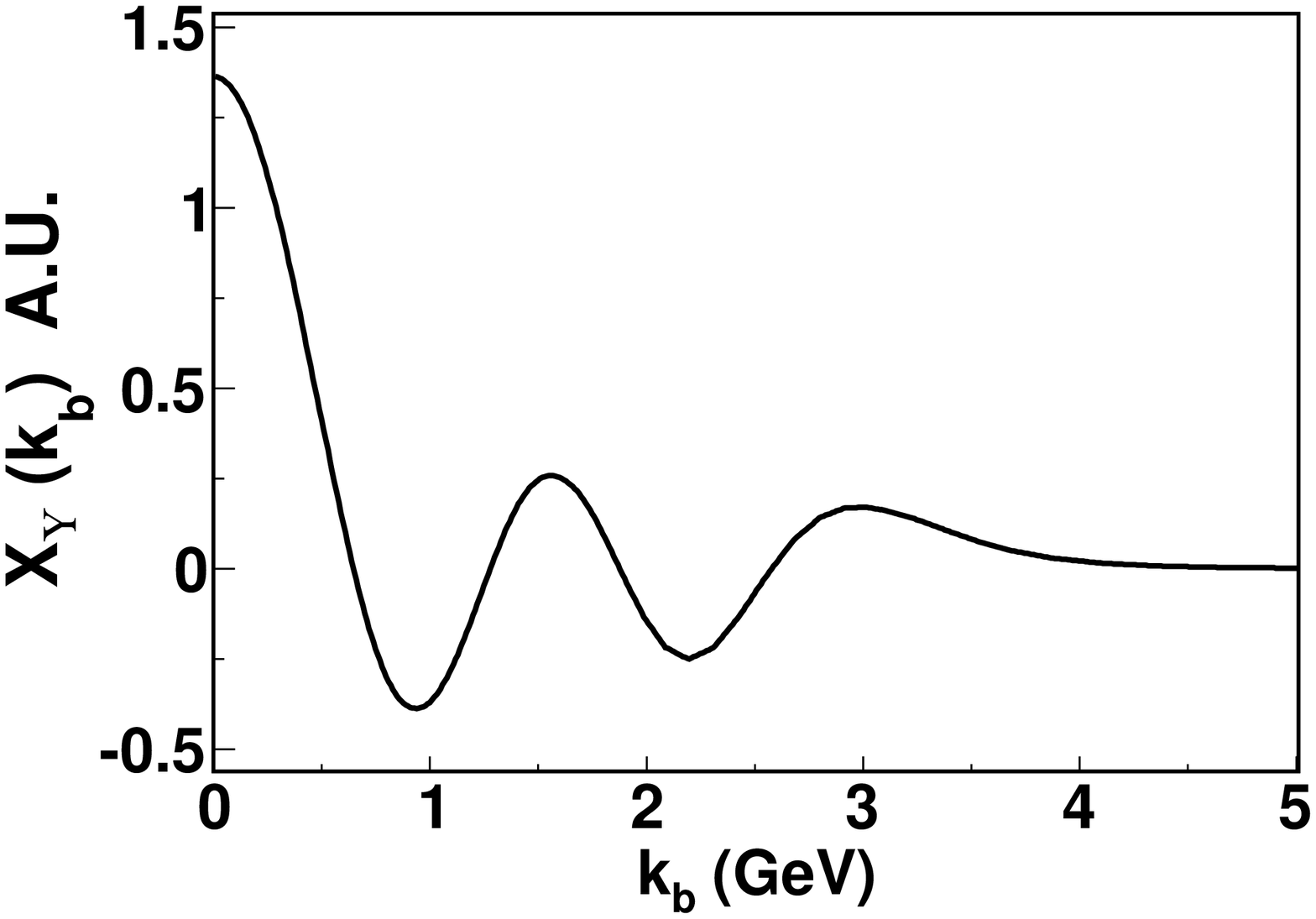}
\includegraphics[width=6.6cm]{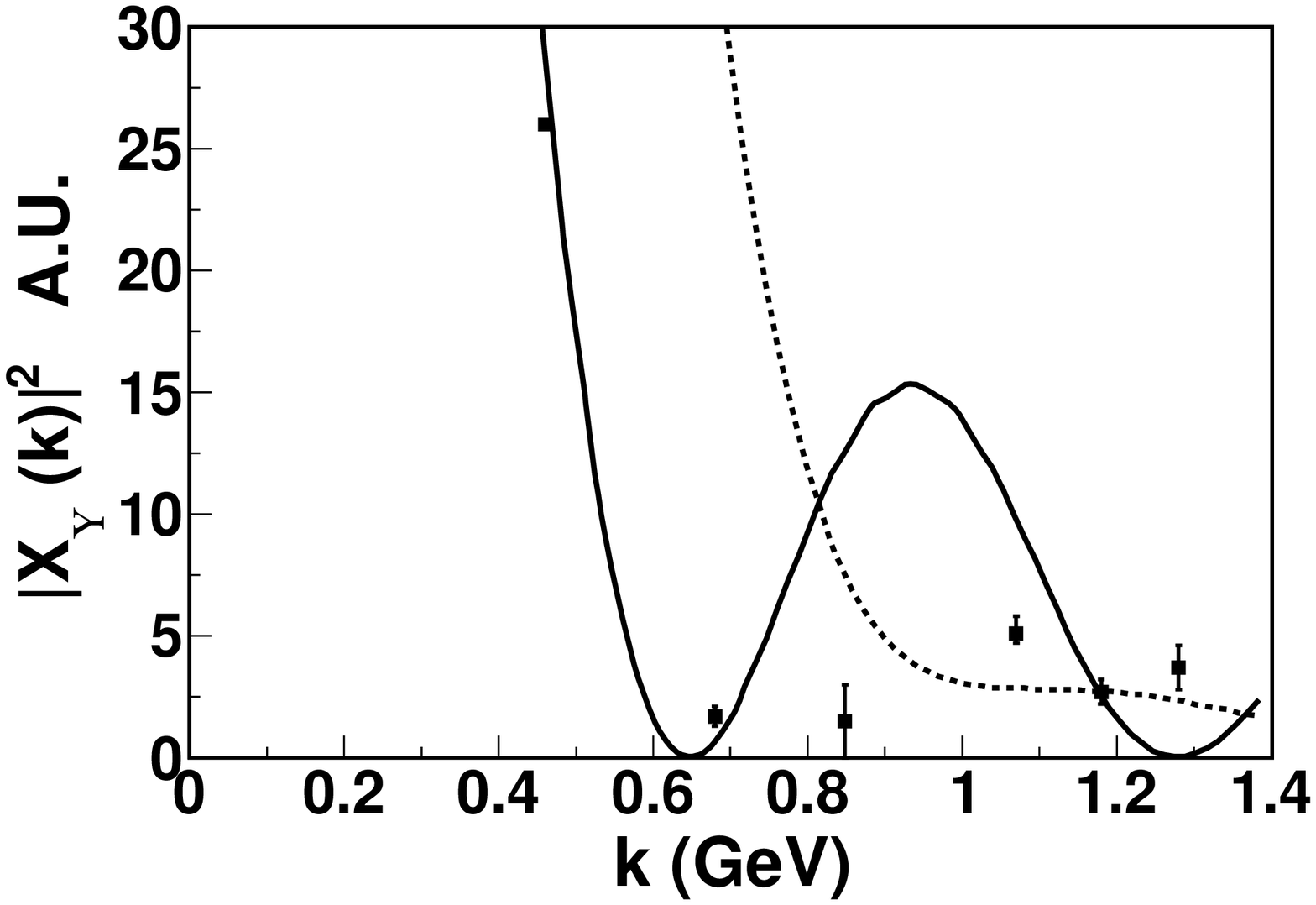}
\caption{Top: Coulomb-gauge model wavefunction of $\Upsilon(5S)$. 
Bottom:
Data points are 2-body decay widths of the $\Upsilon(10860)$ \cite{Drutskoy:2009ci} (Belle collaboration), plotted as a function of the center of mass momentum for the decay products. From right to left, $B\bar{B}$, $B\bar{B}^*+B^*\bar{B}$, $B^*\bar{B}^*$, $B_s\bar{B}_s$, $B_s\bar{B}^*_s+B^*_s\bar{B}_s$, $B^*_s\bar{B}^*_s$. We have divided the experimental data by known spin, isospin and phase-space factors so it can be directly compared to the square of the top panel wavefunction (solid line) or to the wavefunction folded with a complete $^3P_0$ decay model (dashed line), that gives us an idea of the errors of order $\Lambda_{\rm QCD}/m_b$ involved in a naive application of the Franck-Condon principle.
\label{fig:wf}
}
\end{centering}
\end{figure}
Characteristic of pure (excited) quarkonia is the presence of Sturm-Liouville zeroes in the momentum distribution. This follows from the Hermiticity of the QCD Hamiltonian~\cite{Christ:1980ku}, that implies orthogonality between different wavefunctions. Fig. \ref{fig:wf} is a realization of $\Upsilon(5S)$ wavefunction (candidate $b\bar{b}$ assignment for $\Upsilon(10860)$) in a model approximation to Coulomb gauge QCD~\cite{LlanesEstrada:2004wr}. One clearly sees the 4 zeroes of the $5S$ excitation, beautifully exploited in other areas such as nuclear physics~\cite{cavedon:1983}.

The advantage of open-flavor decays (experimentally more difficult to reconstruct than transitions to ground-state $\Upsilon$) is that the velocity $v_b$ of the initial-state quark approximately equals $v_B$, the velocity of the final state meson. 
This is because the final $B$-meson momentum $k$ in a transition driven by gluons and light quarks satisfies
$k=k_b+\mathcal{O}(\Lambda_{\rm QCD})$, which is the rationale behind effective-theory analysis such as Heavy Quark Effective Theory (HQET)~\cite{Isgur:1989vq,Manohar:2000dt}.

\newpage

The 2-body data $\Upsilon(5S) \rightarrow B^{(*)}_{(s)} \ov{B}^{(*)}_{(s)}$
consists of the partial 2-body decay widths. To obtain them from the decay amplitude one integrates over phase space
\be 
d \Gamma = \frac{ (2 \pi)^4 \delta^{(4)} (P-k-k')}{2 M_{\Upsilon}} | \mathcal{M} |^2 \frac{d^3k}{(2\pi)^3 2E_k } \frac{d^3k'}{(2\pi)^3 2E_{k'}}
\ee
where $P$, $k$ and $k'$ are the 4-momenta of the $\Upsilon$, $B$ and $\ov{B}$, respectively. Integration yields the well-known
\be 
\Gamma (k) = \frac{1}{8\pi} \frac{1}{M_{\Upsilon}} \frac{k^2}{E_k E_{k'}} | \mathcal{M} |^2\ . 
\ee
where the modulus of the B-meson momentum $k$ (in the CMS) is fixed by kinematics to
K\"allen's function
$
k  = \frac{1}{2 M_{\Upsilon}} \lambda^{1/2} (M^2_{\Upsilon},M^2,M'^2)$,
with $M,M'$ either of ${M_B,M_{B^*},M_{B_s},M_{B^*_s}}$.
Thus each 2-body decay probes the $k_b$ distribution in the initial state only at one point (blurred by $O(\Lambda_{\rm QCD})$).
But since there are six allowed decay channels, we have access to six different momenta~\cite{Hwang:2008kb}, allowing the suggestive mapping of the $\Upsilon(10860)$ $k_b$ distribution from experimental data in Fig.~\ref{fig:wf} (ignore for now the calculated curves). Automatically one understands the puzzling feature that the decay width to $B_s^{*}\bar{B}_s^{*}$  ($19 \textrm{ MeV} $) is a factor 12 times the width to $B_s\bar{B}_s^{*}+B_s^{*}\bar{B}_s$ ($1.6 \textrm{ MeV}$), instead of the $(7/4)$ ratio expected from spin counting alone~\cite{De Rujula:1976zg}. We can now understand that the second decay just happens to fall near the first Sturm-Liouville zero of the $5S$-wavefunction in the initial state, whereas the first is well into the wavefunction maximum. Thus the anomalously large strange-vector-vector decay finds a simple explanation in terms of the structure of the parent hadron.

If one wishes better resolution of the $k_b$ distribution inside $\Upsilon(10860)$, resort needs to be made to 3-body decays, from which $B^{(*)}\bar{B}^{(*)}\pi$, first analyzed in~\cite{Lellouch:1992bq}, is the more promising given that $\Upsilon(5S)$ is not so high above open--flavor threshold. In these decays the pion carries off a momentum $\mathbf{p}$ that balances momentum conservation. Then if $k'$ (the $\bar{B}^{(*)}$ momentum) is constrained by energy conservation, $k$ (the $B^{(*)}$ momentum)
is a free variable within its phase--space bounds. The partial decay width as a function of this $k$ then allows access to $k_b$ in the initial state.
Performing immediate angular integrals, and leaving  $k\simeq k_b$ free, it is easy to show that
\be 
\frac{d \Gamma(k)}{dk} = \frac{k}{64 \pi^3 M_{\Upsilon} E_k} \int_0^{k'_{max}} dk' \frac{k'}{E_{k'}} | \mathcal{M} | ^2 (k,k',x'), 
\ee
where $k'_{max}$ is the maximum value of the $\bar{B}$-meson momentum given by the kinematics and $x'$ is the polar angle cosine of $\mathbf{k'}$, fixed by energy conservation to \\
$2 k k' x' = (M_{\Upsilon} -E_k -E_{k'})^2-(M_{p}^2 +k^2+k'^2)$.
\begin{figure}[htbp]
\begin{centering}
\includegraphics[width=8cm]{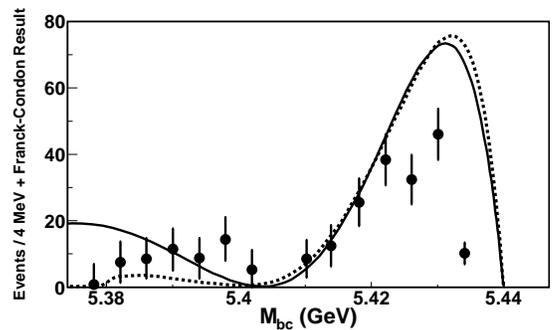}
\caption{Dashed line: computed momentum distribution of the $B^{(*)}$-meson in three-body $\Upsilon(5S)\to B^{(*)}\bar{B}^{(*)}\pi$ non-strange decays (summed over all $B/B^*$ spins). The variable $M_{bc}$ is related with the $B$ meson momentum as $M_{bc} \equiv \sqrt{(M_{\Upsilon}/2)^2-k^2}$.The momentum of the $b$ quark in the parent $\Upsilon(5S)$ is given by $k_b=k+O(\Lambda_{\rm QCD})$. Thus, except at very low $k$, this distribution probes the confined quark degrees of freedom in the initial state. The dip in $k$ at around $600\textrm{ MeV}$ (or $M_{bc} = 5.4$ GeV in the Figure) is caused by the first Sturm-Liouville zero in the $\Upsilon(5S)$ $k_b$ distribution (solid) folded with 3-body phase space and is correlated with the small $B_s\bar{B}_s^*$ decay width (see Fig.~\ref{fig:wf}). This zero is indeed visible in the recent data (Fig.~2b from \cite{bellenew}).
\label{fig:3body}
}
\end{centering}
\end{figure}
If the $\Upsilon(5S)$ has a significant $b\bar{b}$ component, the decay can be interpreted at the quark level as $b\bar{b}\to (b+{\rm light})(\bar{b}+{\rm light})({\rm light})$. Then the Franck-Condon principle (up to corrections $\Lambda_{\rm QCD}/m_b$ that we can estimate within a model) states that the velocity distribution of the final--state $B$ meson is the same as the velocity distribution of the $b$ quark in the initial $\Upsilon(5S)$ state, thus providing us with a window to its structure. In Fig. \ref{fig:3body}
we show the calculated momentum distribution of the $B$ meson in these decays (this is directly measurable by Belle; to obtain a velocity distribution by eye inspection, note $v_B\simeq p_B/m_B$ is quite a good approximation at these energies).
   A dip is clearly visible at around $M_{bc}=5.4\textrm{ GeV}$ (or $k = 600$ MeV). This is a consequence of the first Sturm-Liouville zero in the $\Upsilon(5S)$ wavefunction and is correlated with the small $B_s\bar{B}_s^*$ decay width, since the momentum for that 2-body decay is also around $600$ MeV. Remarkably, structural insight from the parent hadron, in terms of confined degrees of freedom, helps to interpret puzzling features of decay data. Only the lowest $k_b\leq \Lambda_{\rm QCD}$ is not accessible (application of the Franck-Condon principle just shows that the heavy quarks remain at rest up to $\mathcal{O}(\Lambda_{\rm QCD}/m_b)$).

In terms of the Franck-Condon principle, the reported  large fraction of 3-body decays
of $\Upsilon(10860)$~\cite{Drutskoy:2009ci}, that has not been quantitatively predicted by theoretical approaches~\cite{Simonov:2008cr}, is easy to understand. 
Indeed, in these decays there are always final-state configurations available whatever the $k_b$ in the initial state might be, since the pion momentum is an additional free variable to ensure conservation laws. (In 2-body decays on the other hand, only one final $k$ is open, so that only small parts of the $k_b$ wavefunction overlap with the decay operator). 

\begin{figure}[htbp]
\begin{centering}
\includegraphics[width=6.77cm]{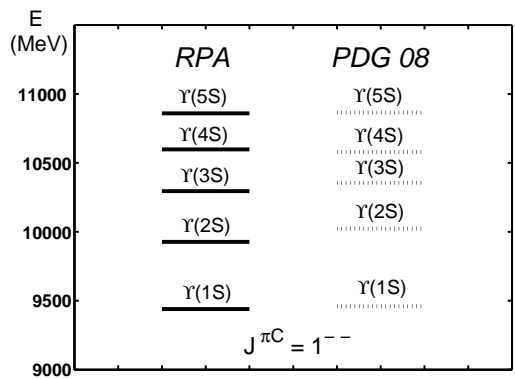}
\caption{$\Upsilon$ spectrum in Cornell-type model of Coulomb-gauge QCD, in agreement with the experimental data~\cite{Amsler:2008zzb}. 
\label{fig:spectrum}
}
\end{centering}
\end{figure}

We now present some calculational detail for the model theory curves provided to illustrate the principle. \\
The structure of the initial $\Upsilon(5S)$ and final states $B$, $\bar{B}^*$, $\pi$, etc. is taken from the Cornell model of Coulomb-gauge QCD based on the Hamiltonian for the quark fields
\begin{eqnarray}  \label{theH}\nonumber
H = - g_s \int d{\bf x} \Psi^\da (x) \alpha  \cd {\bf A}(x) 
\Psi (x)\\ \nonumber
  +
\int d {\bf x} \Psi ^{\dagger} _q
({\bf x}) (-i \alpha\cd \nabla + \beta m_q )
\Psi_q({\bf x}) \\ 
- \frac{1}{2} \int d {\bf x} d {\bf y}
\rho^a({\bf x})V(\arrowvert {\bf x} -
{\bf y} \arrowvert) \rho^a({\bf y}) \ . 
\end{eqnarray}
The $\alpha  \cd {\bf A}$ term in the first line is reduced to a transverse potential between quark sources (of Yukawa type) as explained in~\cite{LlanesEstrada:2004wr} and the Cornell (Coulomb+linear potential) is used for the Coulomb static kernel $V$, a model philosophy  consistent with lattice~\cite{Bali:1997am} results for very heavy quarks.
This relativistic many-body Hamiltonian is treated in BCS+TDA approximations to yield the dressed $b\bar{q}$ or $b\bar{b}$ canonical wavefunctions for pseudoscalar
\be 
\mathcal{X}_{k, \ i j}^{B \ ab} (|\mathbf{q}|)= Y_{00} (\hat{\mathbf{q}}) \left( \frac{i \sigma_2}{\sqrt{2}} \right)_{ij} X_k^B(|\mathbf{q}|) \frac{\delta^{ab}}{\sqrt{3}} 
\ee
and  vector mesons (with spin index $\alpha=1,2,3$).
\be 
\mathcal{X}_{p, \ i j}^{\Upsilon \ ab, \ \alpha} (|\mathbf{q}|)= Y_{00} (\hat{\mathbf{q}}) \left( \frac{i \sigma^{\alpha} \sigma_2}{\sqrt{2}} \right)_{ij} X_p^{\Upsilon}(|\mathbf{q}|) \frac{\delta^{ab}}{\sqrt{3}} \ .
\ee
(We have also computed the D-wave component and found it to be totally negligible in bottomonium). We further neglect the small model difference between the radial $B$ and $B^*$ wavefunctions (which is again supported by HQET for large $m_b$). \\
The values of the model parameters are now given, gluon mass $m_g=0.89$ GeV (string tension, $\sigma$ is dependent as per the cited work of \cite{LlanesEstrada:2004wr}), strong Colomb coupling $g_s=1.41$ and heavy--quark mass $m_b=3.73$ GeV. These parameters yield good agreement with experimental data on the $\Upsilon$ $b\bar{b}$ spectrum as sketched in Figure \ref{fig:spectrum}. 

As for the decay mechanism, a complete QCD-rescattering calculation is beyond our ability, so we supplement the model with the time-tested $^3P_0$ prescription~\cite{barnesswanson}. In this framework, light quark creation proceeds via a chiral--symmetry breaking interaction $H_I$ proportional to ${\bf \sigma}\cdot {\bf p}$.
Dyson--Schwinger studies~\cite{Alkofer:2008tt} have shown that the quark--gluon vertex features strong scalar contributions only up to momenta of order $\Lambda_{\rm QCD}$, where the light quark running mass function $M(k)$ falls steeply like a power--law towards the current--quark mass $m_q$.
Thus we take the $^3P_0$ vertex proportional to the BCS mass--gap $M(k)/E(k)$ generated by Eq.~(\ref{theH}),
that signals spontaneous chiral symmetry breaking, with a proportionality constant $c$ serving as decay--model parameter.

The Feynman amplitude ([mass dimension]=1 for the 2-body decay) parametrizes the S-matrix ($ \mathcal{S} \equiv 1 + i T $), 
\be 
\langle k k' | T | P \rangle = (2 \pi)^4 \delta^{(4)} (P-k-k') \mathcal{M} \ .
\ee
The Born series $i T =- i \int dt H_{I} + \cdots$ connects it then to
the $^3P_0$ decay--model Hamiltonian
\ba \label{hi}
 - \int dt \langle k k' | H_I | P \rangle = (2\pi)^4 \delta^{(4)} (P-k-k') \mathcal{M} 
\\ 
H_I\equiv -c \int \frac{ d^3 q}{(2\pi)^3} \frac{M(q)}{E(q)} \left( {\bf \sigma} \cdot {\bf \hat{q}} i \sigma_2 \right) B_q^{\dag} D_{-q}^{\dag}. 
\ea
The time independence of $H_I$ leaves, as customary, the decay rate
$
\lim_{T \rightarrow \infty} \int_{-T}^T \langle k k' | H_I | P \rangle dt = \delta(0) \langle k k' | H_I | P \rangle 
$.
\begin{figure}[htbp]
\begin{centering}
\includegraphics[width=8cm]{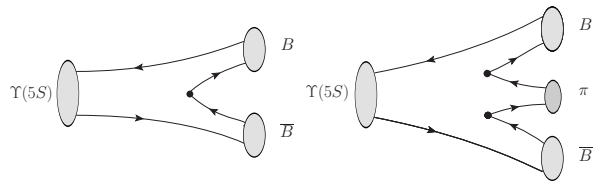}
\caption{Model $\Upsilon$ 2- and 3-body open-flavor decays through $^3P_0$ vertices. In the limit $\Lambda_{\rm QCD}/M_b \to 0$ they respect the Franck-Condon principle that QCD should also obey, and the momenta of the heavy quark lines flowing from the left is close to the external momenta of the decay mesons on the right.
\label{fig:23decay}
}
\end{centering}
\end{figure}
The decay amplitude is then depicted in Fig.~\ref{fig:23decay} (left).
Its computation requires the matrix element in Eq.~(\ref{hi}),
\ba \label{amplitucita}  
\langle k k' | H_I | P \rangle = c \delta^{(3)} ({\bf k} + {\bf k'}) \int d^3 {\bf q}   \frac{M(|{\bf k}- {\bf q}|)}{E(|{\bf k}- {\bf q}|)} 
\ \ \ \\ 
\nonumber \times Tr \left\{ \left[ \mathcal{X}_k \left( \left| \frac{{\bf k}}{2} - {\bf q} \right| \right) \right]^2
\mathcal{X}_{P} \left( | {\bf q} | \right) \
\left( {\bf \sigma}  \cdot \frac{ {\bf k}-{\bf q} }{|{\bf k}-{\bf q}| } \right) i \sigma_2 \right\}.
\ea

The Franck-Condon principle is patent in this expression; 
if the decay vertex $M(|{\bf k}- {\bf q}|)$ is small for $|{\bf k}- {\bf q}|\gg\Lambda_{\rm QCD}\simeq 250 \textrm{ MeV}$,  the heavy--quark momentum ${\bf k} - 2{\bf q}$, which is twice the argument of the $B$-meson wavefunction $\mathcal{X}_k $ in Eq.~(\ref{amplitucita}), becomes the external meson momentum ${\bf k}$ up to $O(\Lambda_{\rm QCD})$. The heavy--quark and meson velocities are then approximately the same as expected. Thus we have a simple model realization of the general principle.

We average over initial and sum over final spin states for unpolarized decays. Particularizing to $B\ov{B}$, we have
\be 
\langle | \mathcal{M}|^2 \rangle = \frac{1}{3} \sum_{\alpha} \mathcal{M}^{* \alpha} \mathcal{M}^{\alpha}. 
\ee
In 2-body channels the spin factorizes from the integral for all three final state spin combinations, leaving the known ratios  1:4:7 for $B\bar{B}:
B\bar{B}^*+B^*\bar{B}:B^*\bar{B}^*$.

The 3-body decays depicted on the right of Fig.~\ref{fig:23decay} are an immediate extension of these formulae, except that
the pion wavefunction requires an additional word of caution. It is computed in the rest frame defined by the Hamiltonian in Eq.~(\ref{theH}) but the velocity of the outgoing pion is relativistic. While rigorous studies of the boost operator in Coulomb QCD are underway~\cite{Rocha:2009xq}, in this work we have limited ourselves to a kinematic Lorentz contraction of the computed wavefunction.

Looking forward to further refinements of this work,
we note that our estimate of the divergence from the Franck-Condon principle is model--based, and perhaps one could directly use effective theory language to parametrize this irreducible error. But it is not currently clear what effective theory one should pursue.

Ground state $b\bar{b}$ is deeply bound respect to the open--flavor decay threshold. Typical momentum transfers are of order $\alpha_s M_b$ and do not vanish asymptotically for large quark masses at fixed $\alpha_s$, the 
appropriate effective theory being NRQCD. 
However, for resonances above threshold such as $\Upsilon(10860)$ some extension of HQET might apply, since  the open flavor $B$ mesons perceive momentum transfers from the vertex to the heavy quark of order the string tension scale, not proportional to $m_b$. This deserves further study. In both cases though, the leading order phenomenon is the equality of heavy quark and heavy meson velocity that we have exploited.

On the phenomenological side, a large number of new resonances in the high-charmonium spectrum, $3.8-4.6\ \textrm{GeV}$, has been unveiled in the last years. These new states, while pending classification named $X(3940)$, $Y(4260)$, $Y(4320)$, ... or generically $X-Y-Z$ resonances, are seen either recoiling against a $J/\psi$ (such as the 3940) or decaying to $J/\psi +n\pi$ with $n=1,2,3\dots$

Unstable resonances pose very challenging problems to lattice gauge theory.
No known effective theory can describe either of all these resonances from first principles without additional model assumptions. The amount of ``hard'' theoretical understanding is then very limited, and the discussion is still at the level of simple models.

Our observation is that some structural information (such as momentum distributions of heavy quarks inside the new states) 
can be obtained from open--flavor decays.
Several examples~\cite{General:2006ed} have been presented of the possible use of the fluorescence analogy to analyze the decays of the new charmonium-like states $XYZ$. 
 A (presumedly) excited $c\bar{c}$ state would feature Sturm-Liouville zeroes in the decay $\psi^*\to D\bar{D}\pi$, by tracking the momentum of the $D$. An exotic meson (hybrid, tetraquark, etc) of equal mass would not present such Sturm-Liouville nodes in its relative $c\bar{c}$ momentum distribution. 
\\ 
Another application consists of constructing an off--planes correlator 
in four--body decays to identify the likeliness of a tetraquark component in the excited meson.\\
Yet a third idea is to use $D\bar{D}$ momentum distributions in certain $B$--meson decays to learn about the charm--sea in hadrons. We look forward to turning some of these possibilities into usable experimental tests.

\begin{acknowledgments}
We thank I.~Scimemi, C. Hanhart, A.~Dobado and J.~Soto for useful conversations, Alexey Drutskoy for insight into the published Belle measurements, and grants FPA
2008-00592, FIS2008-01323 plus FPU graduate support for J.T. (Spain), and 227431, HadronPhysics2 (EU).
\end{acknowledgments}

\end{document}